\documentclass[preprint]{aastex} 
\input{epsf}
\newcommand{\mpc}{h^{-1}{\rm Mpc}}

\newcommand{\gsim}{\mathrel{\hbox{\rlap{\hbox{\lower4pt\hbox{$\sim$}}}\hbox{$>$}
}}}
\newcommand{\lsim}{\mathrel{\hbox{\rlap{\hbox{\lower4pt\hbox{$\sim$}}}\hbox{$<$}
}}}

\shortauthors{Rojas et al.}
\shorttitle{Spectroscopic Properties of Void Galaxies}

\begin{document}
\title{Spectroscopic Properties of Void Galaxies in the Sloan Digital
Sky Survey} \author{Randall R. Rojas\altaffilmark{1}, Michael S. Vogeley\altaffilmark{1},
Fiona Hoyle\altaffilmark{1}, \& Jon Brinkmann\altaffilmark{2} \\
\email{rrojas@mercury.physics.drexel.edu, vogeley@drexel.edu,
fiona.hoyle@drexel.edu }}

\altaffiltext{1}{Department of Physics, Drexel University,
3141 Chestnut Street, Philadelphia, PA 19104}
\altaffiltext{2}{Apache Point Observatory, P. O. Box 59, Sunspot, NM 88349-0059}
 
\begin{abstract}

We study the spectroscopic properties of a sample of $10^3$ void
galaxies from the Sloan Digital Sky Survey (SDSS) and compare these
with the properties of galaxies in higher density regions (wall
galaxies). This sample of void galaxies covers the range of absolute
magnitude from $M_r=-13.5$ to $M_r=-22.5$ in regions with density
contrast $\delta\rho /\rho<-0.6$. In this paper we compare the
equivalent widths of H$\alpha$, [OII], [NII], H$\beta$, and [OIII] of
void and wall galaxies of similar luminosities and find that void
galaxies have larger values, indicating that they are still forming
stars at a high rate. A comparison of the Balmer break, as measured by
the parameter D$_{n}(4000)$, reveals that void galaxies have younger
stellar populations than wall galaxies. Using standard techniques, we
estimate H$\alpha$ and [OII] star formation rates of the void and wall
galaxies and along with estimates of the stellar masses, we compute
specific star formation rates. In most cases, we find that void
galaxies have similar SFRs to wall galaxies but they are fainter and
smaller mass. This means that, consistent with the EWs, void galaxies
have higher specific star formation rates than wall galaxies.
\end{abstract}


\section{Introduction}
\label{sec:intro}

Studying the variation within environment of galaxy properties is
fundamental to our understanding of galaxy formation and evolution.
Dressler (1980) characterized the morphology-density
relation -- the tendency for elliptical galaxies to reside in
clusters and spirals in the field. This has been subsequently
confirmed by many different groups (e.g.  Postman \& Geller 1984;
Zabludoff \& Mulchaey 1998; Hashimoto \& Oemler 1999; Tran et
al. 2001; Dominguez et al 2001; Treu et al. 2003). Recent studies
have attempted to extend the study of this environmental dependence into
increasingly underdense regions. In a series of papers, we seek to
examine if galaxies that reside in the most underdense regions,
referred to as void galaxies, are fundamentally different from galaxies
in denser environments.

Many studies have also shown that there is a strong dependence of star
formation on environment. High star formation rates (SFR) are usually
found in low density environments while lower SFRs are more commonly
seen in higher density regions. (Dressler, Thompson \& Shectman 1985;
Couch \& Sharples 1987; Balogh et al. 1997, 1998, 1999, 2002;
Hashimoto et al. 1998; Poggianti et al. 1999; Couch et al. 2001;
Solanes et al. 2001; Lewis at al. 2002a; Gomez et al. 2002).

However, only recently was it possible to show that there is a strong
dependence of SFR on local galaxy density in large enough samples to
include a fair mix of morphological types. Gomez et al. (2002) showed
that there is a decrease in SFR from galaxies in the field to galaxies
in denser environments (e.g. in rich clusters). Independently, Lewis
et al. (2002a) found the same strong correlation between environment
and SFRs in the 2dF Galaxy Redshift Survey. Hashimoto et al. (1998)
noticed that regardless of the concentration index, galaxies in
clusters have on average lower SFRs than field galaxies in the Las
Campanas Redshift Survey. As well as looking at SFRs, some groups have
considered the strength of H$\alpha$ emission and consistently find
that H$\alpha$ emission has a strong dependence on local density
(Balogh et al. 2004; Tanaka et al. 2003). Tremendous efforts in
extrapolating the {\it density-sfr} relation down to lower densities
have met with success. However, none have been able to probe the very
lowest densities.

Void galaxies have been studied photometrically by Grogin \& Geller
(1999) and by Rojas et al. (2004). They demonstrated that there are
significant differences between void and non-void galaxies, with
void galaxies being bluer in color and having morphologies that more
closely represented late-type galaxies. However, they found that
this was not just an extension of the morphology-density relation
as late-type galaxies in low density environments were bluer than
late-type galaxies in higher density environments and the same
help for early-type galaxies i.e. density plays a role in the
properties of the same type if galaxies rather than just changing
the relative mix of type of galaxy. 

In this paper, we focus on the spectroscopic properties of void
galaxies to examine the role of density on the strength of the
equivalent widths of various lines and on star formation rates. Using
a sample of 150 galaxies with $\delta \rho / \rho < 0$, Grogin \&
Geller (2000) showed that the EW(H$\alpha$)-density relation is most
noticeable in either the lowest or highest density environments i.e
the void galaxies had large EW(H$\alpha$) and the galaxies in richer
environments had smaller values of EW(H$\alpha$). We extend this work
by using a sample of $10^{3}$ void galaxies with density contrast
$\delta \rho/ \rho<-0.6$ that spans a broad range of absolute
magnitude from $M_{r}=-13.5$ to $M_{r}=-22.5$. We use this sample to
quantitatively extend the study of star formation history down to the
lowest densities possible and compare the spectroscopic properties of
our void galaxies with non-void galaxies of similar luminosities and
surface brightness profiles. These comparisons are crucial in order to
constrain models of galaxy formation and evolution.

In Section \ref{sec:data} we discuss the data sample. In Section
\ref{sec:spec} and \ref{sec:specdervied} we present the results of the
various spectroscopic parameters. We compare the results with those
from other groups in Section \ref{sec:disc}. Finally, in Section
\ref{sec:conc} we present our conclusions.

\section{The Data}
\label{sec:data}
We identify a a sample of 10$^3$ void galaxies from early data
available from the SDSS. The SDSS is a wide-field photometric and
spectroscopic survey that will cover approximately $10^{4}$ square
degrees. CCD imaging of 10$^8$ galaxies in five colors and follow-up
spectroscopy of 10$^6$ galaxies with $r<17.77$ will be obtained. York
et al. (2000) provides an overview of the SDSS and Stoughton et
al. (2002) describes the early data release (EDR) and details about
the photometric and spectroscopic measurements and Abazajian et
al. (2003, 2004) describe the first (DR1) and second (DR2) data
releases.  Technical articles providing details of the SDSS include
descriptions of the photometric camera (Gunn et al. 1998), photometric
analysis (Lupton et al. 2002), the photometric system (Fukugita et
al. 1996; Smith et al. 2002), the photometric monitor (Hogg et
al. 2001), astrometric calibration (Pier et al. 2003), selection of
the galaxy spectroscopic samples (Strauss et al. 2002; Eisenstein et
al. 2001), and spectroscopic tiling (Blanton et al. 2003a). A
thorough analysis of possible systematic uncertainties in the galaxy
samples is described in Scranton et al. (2002). All the galaxies are
k-corrected according to Blanton et al. (2003b) and we assume a
$\Omega_{\rm m}, \Omega_{\Lambda}$ = 0.3, 0.7 cosmology and Hubble's
constant, $h = H_o/ 100$ km s$^{-1}$Mpc$^{-1}$ throughout.

In this paper, we consider the spectral properties of galaxies in
different environments. The SDSS spectra cover the wavelength range
3900-9100\AA, obtained using a double blue/red spectrograph with the
beam split at 6150\AA\ by a dichroic. The data have spectral
resolution ranging from 1800 to 2100 (i.e. $\lambda / \delta \lambda
\approx 2000$), so at 4000\AA\ the resolution is $\sim$2\AA. The
signal-to-noise ratio, S/N, is typically 13 per pixel.

Void galaxies are drawn from a sample referred to as {\tt sample10}
(Blanton et al. 2002). This sample covers nearly 2000 deg$^2$ and
contains 155,126 galaxies. We use a nearest neighbor analysis,
described in more detail below, to find galaxies that reside in region
of density contrast $\delta \rho / \rho < -0.6$ as measured on a scale
of 7$h^{-1}$Mpc, which we label void galaxies. Galaxies with larger
vales of $\delta \rho / \rho$ are referred to as wall galaxies.  Full
details of void galaxy selection are described in Rojas et
al. (2004). Here we provide a brief overview, as follows.  First, a
volume-limited sample of relatively bright galaxies is constructed to
define the density field that traces the distribution of voids.  This
volume-limited sample extends to maximum redshift z$_{\rm max}=0.089$.
We identify void galaxies from the flux limited sample, also truncated
at z$=0.089$. We discard galaxies that lie close to the edge of the
survey, because it is impossible to tell if a galaxy is a void galaxy
if its neighbors could not yet have been observed.  For each of the
remaining galaxies in the flux-limited sample, we measure the distance
to third nearest neighbor the volume-limited catalog.  A galaxy with
fewer than three neighbors within a sphere of radius 7$h^{-1}$Mpc is
flagged as a void galaxy. Galaxies with more than 3 neighbors we label
as wall galaxies. This procedure yields a sample of 1,010 void
galaxies and 12,732 wall galaxies. These void and wall galaxies span a
redshift range $0.034\lsim z < 0.089$ and a range of absolute
magnitudes from $-22<$M$_{\rm r}<-17.77$. This sample is referred to
as the ``distant'' sample. This sample is also split in absolute
magnitude at $M_{r}=-19.5$ into bright [WGD\_b, VGD\_b]
($M_{r}\leq -19.5$, b=bright, W=wall, V=void) and faint [WGD\_f,
VGD\_f] ($M_{r}> -19.5$, f=faint) sub-samples respectively.

We construct a sample of fainter and more nearby void galaxies -- the
``nearby'' sample -- by using the wider angle Updated Zwicky Catalog
(Falco et al. 1999) and the Southern Sky Redshift Survey (da Costa et
al. 1998) to trace the distribution of local voids (see figure 1. in
Rojas et al. 2004), at distances where the slices of available SDSS
scans (which extend over roughly 2.5 degrees in declination) are too
narrow to accurately map the large-scale structure.  We construct
volume-limited samples of the UZC and SSRS2 to match the density of
objects used as tracers for the ``distant'' sample of SDSS galaxies.
These nearby samples extend to maximum redshift z$_{\rm max}=0.025$.
We apply the same nearest neighbor analysis and identify a further 194
void galaxies and 2,256 wall galaxies from a flux-limited sample of
SDSS galaxies. The absolute magnitudes of these galaxies lie in the
range $-19.7<$M$_{\rm r}<-13$. We also split the Nearby sample into
bright [WGN\_b, VGN\_b] ($M_{r}\leq -17$, b=bright) and faint [WGN\_f,
VGN\_f] ($M_{r}> -17$, f=faint) sub-samples.

\section{Measured Spectroscopic Properties}
\label{sec:spec}

In this paper we study the spectroscopic differences between void and
wall galaxies. We compare their H$\alpha$, [OII] $\lambda3727$,
H$\beta$, [NII] $\lambda6583$, [OIII] $\lambda5007$ equivalent widths
(EWs), stellar masses, 4000 {\AA} Balmer break (ratio of the average
flux density in the bands 4050-4250 and 3750-3950 {\AA}, D$_{n}(4000)$;
Bruzual 1983).

We choose these quantities to examine a broad range of features in
the spectra of our galaxies and star formation on different time
scales. For example, the H$\alpha$ line probes star formation on
timescales of $\sim10^{7}$ years (lifetime of an HII region)
whereas the 4000 {\AA} break probes timescales of $t\sim1$ Gyr.

These emission lines are sensitive to a range of properties of star
formation regions. EW(H$\alpha$) measures the ratio of flux from
recent star formation (mainly flux from UV photons i.e, $\lambda<912$
{\AA}) to the integrated past star formation (predominantly flux from
old stellar populations) whereas the [OII] forbidden line is due to UV
radiation ($\lambda<730$ {\AA}) produced by massive stars that ionize
heavier elements. The [OII] line's dependence on excitation and oxygen
abundance makes it a somewhat less reliable measure of current star
formation (Koo and Kron 1992). Note that [OII] cannot be detected in
90\% of the nearby galaxies because it can only be measured by the
SDSS at a redshift of $z\approx 0.07$. No [OII] results are given for
the nearby sample.

EW(H$\alpha$) is typically double that of [OII] but the [OII] line
is less affected by dust than H$\alpha$. In both cases, large
values of EW suggest high star formation. Large H$\alpha$ equivalent
widths are directly related to high star formation rates, bluer colors
and fainter absolute magnitudes (Kennicutt \& Kent 1983). Strong
correlations between EW([OII]) and galaxy colors have been shown to
exist for various galaxy types and redshifts (Huchra 1977; Peterson
et al. 1986; Broadhurst, Ellis, \& Shanks 1988; Colless et al. 1990;
Broadhurst, Ellis, \& Glazebrook 1992; Koo \& Kron 1992).

We use the widths of the Gaussian fit to the respective lines as
measured by the SDSS spectral pipeline ({\tt spectro1d}; SubbaRao et
al, 2004). For the 4000 {\AA} break strength we use, where possible,
Kauffmann et al's. (2003; K03) estimates of this index
(D$_{n}$4000). They use the ratio of the average flux density in
narrow continuum bands (3850-3950 and 4000-4100 {\AA}) as suggested by
Balogh et al. (1999). D$_{n}(4000)$ is an excellent indicator of
past star formation on timescales of $t\sim1$Gyr and is also
insensitive to dust attenuation. For hot stars where elements are
multiply ionized the opacity decreases. Therefore, hot, young stars
(e.g., O and B) have lower amplitudes of D$_{n}(4000)$ than cooler,
older stars (e.g., K and M; see Figure 3, Bruzual 1983).  Since
late-type galaxies are mainly composed of young stars, their 4000
{\AA} break is smaller than for early-type galaxies which are
primarily composed of older stellar populations.

We compute the means and the errors on the mean of the distributions
of EW's and $D_n(4000)$'s of void and wall galaxies. We also use the {\it
Kolmogorov-Smirnov} (KS) test to examine whether the void and wall
galaxies could be drawn from the same parent population based on the
spectroscopic properties under consideration.

Tables 1 and 2 summarize the results of these tests for the nearby
and distant samples respectively. We present results for the whole
sample, as well as the samples split by absolute magnitude. The
results are considered in detail below.

\subsection{Equivalent Widths}
\label{sec:eqw}

\begin{figure}
\begin{centering}
\begin{tabular}{c}
{\epsfxsize=11truecm \epsfysize=11truecm \epsfbox[35 170 550 675 ]{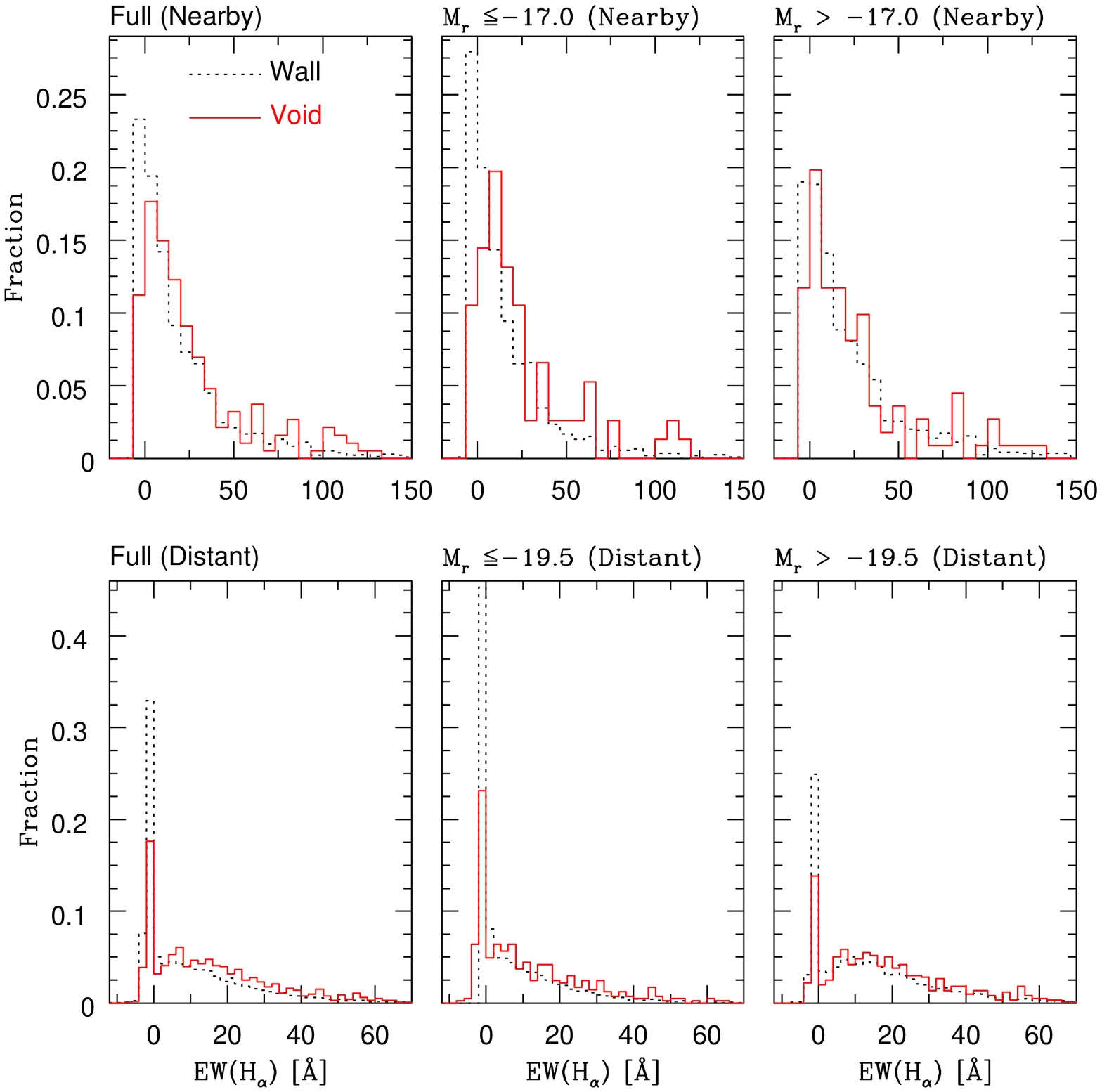}}
\end{tabular}
\caption{Distribution of H$\alpha$ equivalent widths. We show the
normalized fraction of void (solid lines) and wall galaxies (dotted
lines) as a function of EW(H$\alpha$). The top row shows the results
for the nearby ($r<72 \mpc$) galaxies, the bottom row shows the
results for the distant ($100\leq r \leq260 \mpc$) galaxies.  The
first, second and third columns are the full, bright and faint
samples. The fraction of galaxies per 6.5{\AA} (nearby) and 2{\AA}
(distant) bin of EW(H$\alpha$) is shown on the {\rm Y}-axis.  The KS
statistic reveals that the distant void galaxy (bright, faint, and
full) and respective wall galaxy samples are very different from one
another, with a probability of $<0.01\%$ that they are drawn from the
same parent population. In the case of the nearby galaxies, only the
faint galaxy distributions (top right panel) have a higher probability
($P\lsim0.37$) of being similar. }
\label{fig:haew}
\end{centering}
\end{figure}

\begin{figure}
\begin{centering}
\begin{tabular}{c}
{\epsfxsize=11truecm \epsfysize=11truecm \epsfbox[35 170 550 675 ]{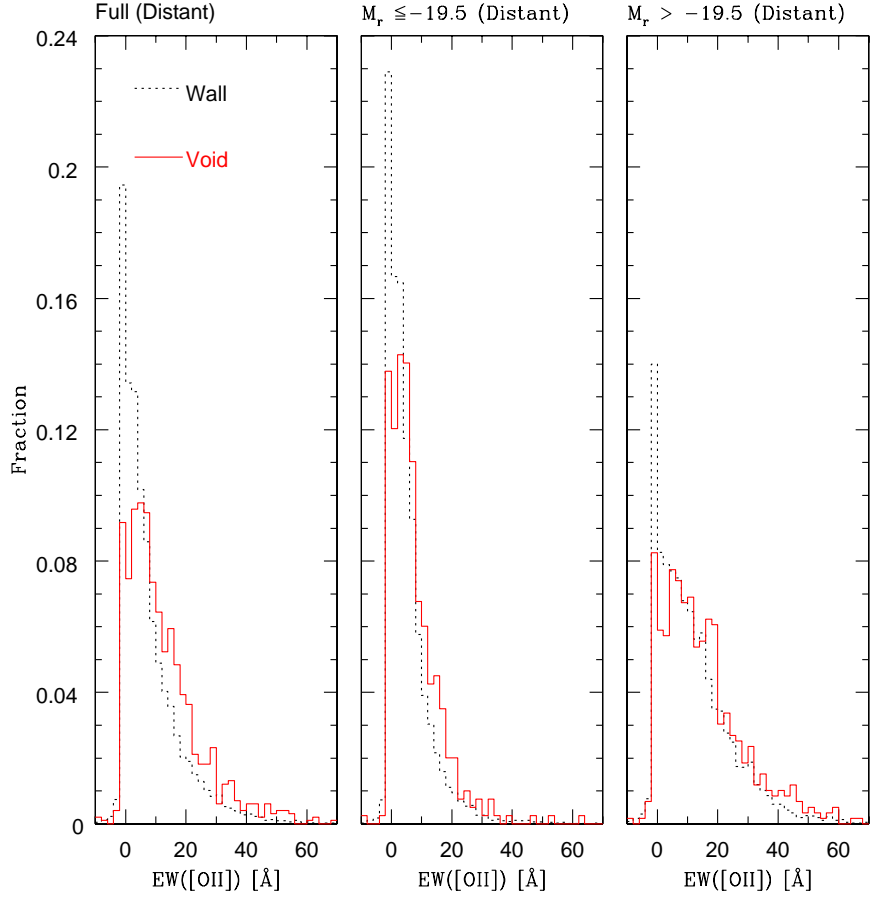}}
\end{tabular}
\caption{Distribution of [OII] equivalent widths. We show the
normalized fraction of void (solid lines) and wall galaxies (dotted
lines) as a function of EW([OII]).  The first, second and third
columns are the full, bright and faint distant samples. The fraction
of galaxies per 2{\AA} bin of EW([OII]) is shown on the {\rm Y}-axis.
The KS statistic reveals that the distant void galaxy (bright, faint,
and full) and respective wall galaxy samples are very different from
one another, with a probability of $<0.01\%$ that they are drawn from
the same parent population. }
\label{fig:o2ew}
\end{centering}
\end{figure}

In Tables 1 and 2 all the values of the EW of the different lines
(H$\alpha$, [OII] (where possible), H$\beta$, [NII] and [OIII]) and
samples (distant, nearby, bright, faint) are given. In Figures
\ref{fig:haew} and \ref{fig:o2ew} we show normalized histograms of
EW(H$\alpha$) and EW([OII]) for the different samples. The solid lines
correspond to the void galaxies and the dotted lines to the wall
galaxies.

Void galaxies have average EW(H$\alpha$) of 19.14$\pm0.68$ in the
distant sample whereas wall galaxies have
EW(H$\alpha$)=11.77$\pm0.16$. The quoted errors are the error on the
means and these will be quoted throughout. In the nearby sample, the
values are higher, EW(H$\alpha$)=35.31$\pm0.26$ (void) and
26.18$\pm0.65$ (wall). The values for EW([OII]) in the distant sample
are 14.26$\pm0.40$ (void) and 9.44$\pm0.09$ (wall). In all the distant
samples, we find that void galaxies have on average larger EWs than
wall galaxies with differences in the means $>5\sigma_{\mu}$. In the
nearby samples, the differences are around 2$\sigma_{\mu}$ due to the
smaller sample size.

The statistical significance of the EW(H$\alpha$) differences between
void and wall galaxies is quantified through a KS test. For the
distant samples it is very unlikely that the void and wall galaxies
are drawn from the same parent population ($P<10^{-4}$). In the case
of the nearby samples only the faint sub-sample shows a larger
probability ($P=0.37$) of being drawn from the same parent population.
We attribute this in part to the the decrease in number of
galaxies in this sample.

A further test that we do is to split the distant galaxies by
morphology. Galaxies with S\'ersic index $<1.8$ have surface
brightness profiles that resemble spirals, where as galaxies with
S\'ersic index $>1.8$ are more like ellipticals (see Rojas et al. 2004
for the justification). In these cases, void galaxies also have
statistically significant larger EWs.

\subsection{D$_{n}({4000})$}

From Tables 1 and 2 we can see that there is a very low probability
($P<10^{-4}$) that the distributions of D$_{n}(4000)$ of void and wall
galaxies (Distant and Nearby) are drawn from the same parent
population.

\begin{figure}
\begin{centering}
\begin{tabular}{c}
{\epsfxsize=11truecm \epsfysize=11truecm \epsfbox[35 170 550 675 ]{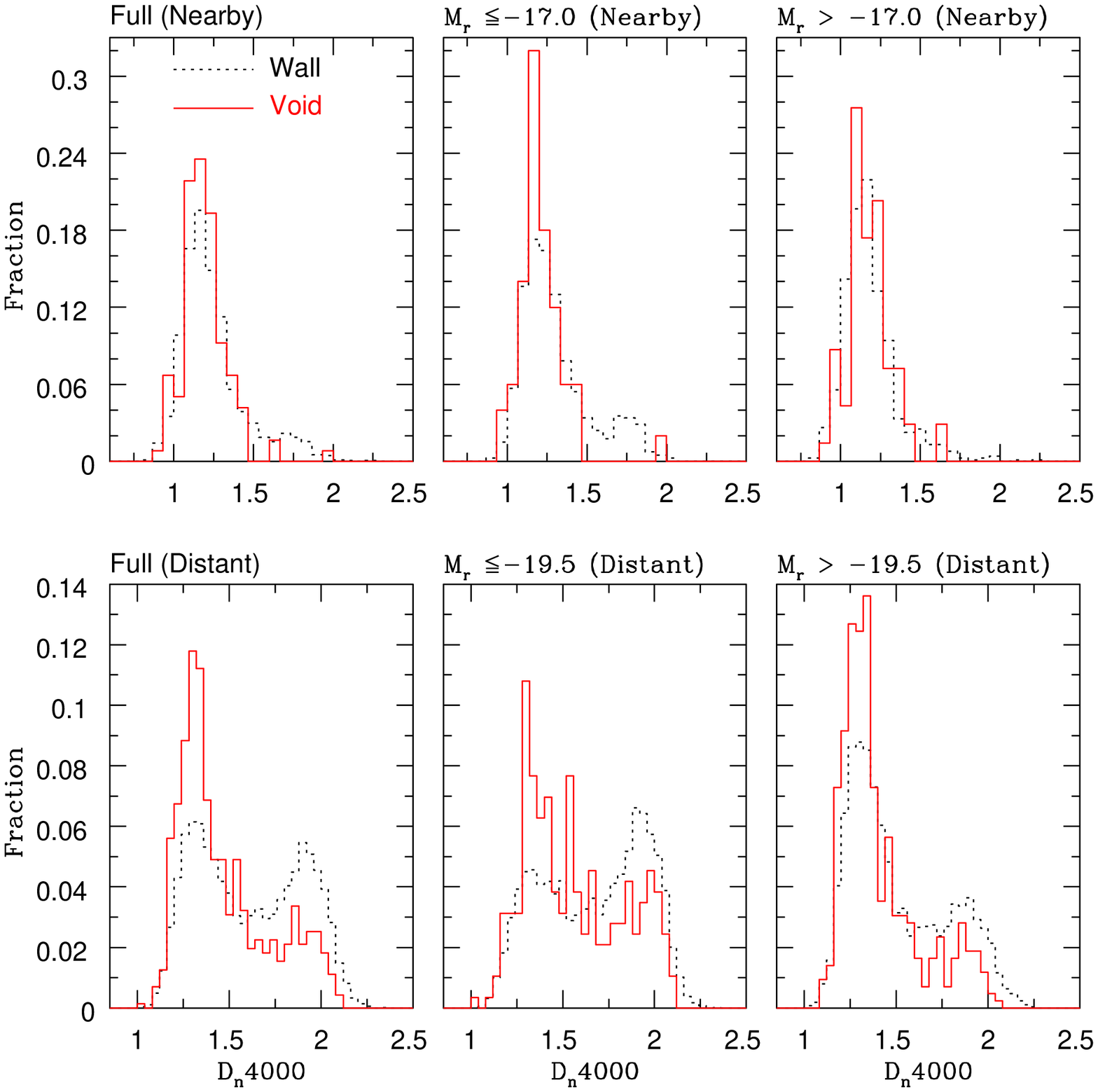}}
\end{tabular}
\caption{Distribution of the 4000 {\AA} Balmer break. We show the
normalized fraction of void (solid lines) and wall galaxies (dotted
lines) as a function of D$_{n}4000$. The top row shows the results for
the nearby ($r<72 \mpc$) galaxies, the bottom row shows the results
for the distant ($100\leq r \leq260 \mpc$) galaxies.  The first,
second and third columns are the full, bright and faint samples. The
fraction of galaxies per 0.067 (nearby) and 0.04 (distant) bin of
D$_{n}4000$ is shown on the {\rm Y}-axis.  The KS statistic reveals
that the distant and nearby void galaxy (bright, faint, and full) and
respective wall galaxy samples are very different from one another,
with a probability of $<0.01\%$ that they are drawn from the same
parent population. }
\label{fig:dn4000}
\end{centering}
\end{figure}

In Figure \ref{fig:dn4000} we show the distributions of D$_{n}(4000)$
for the void and wall galaxy samples. We can observe two noticeable
features: (1) all samples of void galaxies have on average smaller
D$_{n}(4000)$ than wall galaxies and (2) there is an obvious deficit
of early-type galaxies in the nearby samples which is observed in the
amplitude of the distributions for D$_{n}(4000)\gsim1.5$. This is
consistent with the results of Rojas et al. (2004), where there were
few galaxies with S\'ersic and concentration indices typical of
early-type galaxies. The lower values of D$_{n}({4000})$ indicate that
there is more recent star formation in the void galaxies.

\section{Derived Spectroscopic Properties}
\label{sec:specdervied}

An alternative way of comparing the spectral properties of void and
wall galaxies is to convert the spectral lines into estimated star
formation rates (SFR). This is discussed below in section \ref{sec:sfr}.
We also wise to compute the {\it specific} star formation rate,
i.e. the SFR divided by the mass of the galaxy. For this, we obviously
require masses and we discuss this first.

\subsection{Stellar Masses}
\label{sec:mass}

\begin{figure}
\begin{centering}
\begin{tabular}{c}
{\epsfxsize=11truecm \epsfysize=11truecm \epsfbox[35 170 550 675 ]{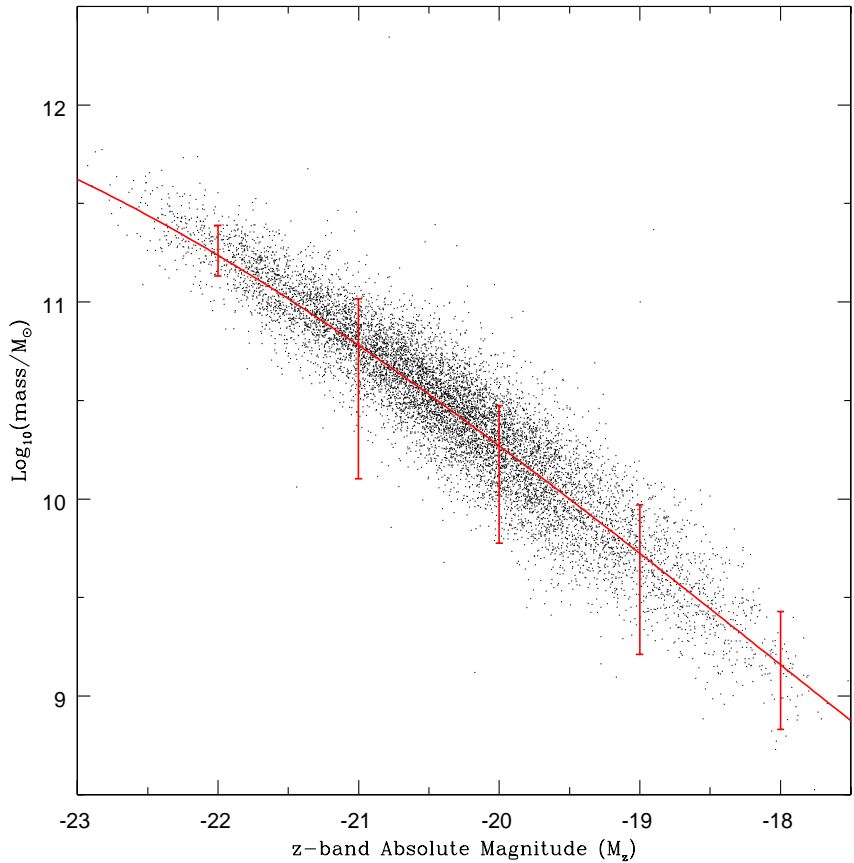}}
\end{tabular}
\caption{Plot of $\rm{Log_{10}(Mass/M_{\odot})}$ vs. $z$-band absolute
magnitude using Kauffmann's et al (2003) stellar masses
(points). Error bars are the $1\sigma$ errors of stellar masses in
bins $\Delta M_{z}=1$ wide. The solid line is a least squares fit. We
use this linear fit to estimate the stellar masses of other galaxies
in our sample using their $z$-band flux.}
\label{fig:fit}
\end{centering}
\end{figure}

The integrated stellar masses for our 70\% of our sample come from
K03's library of stellar masses. They estimate the stellar masses by
taking the product of the model predicted $z$-band
stellar-mass-to-light ratio (M/L)$_{z}$ and the respective
dust-corrected luminosity of the galaxy. The model used by K03 to
predict the (M/L)$_{z}$ and stellar masses is based on median
likelihood estimates from their library of Monte Carlo realizations of
different star formation histories. 30\% of the galaxies are missing
from K03 as they took galaxies from DR1 (Abazajian 2003) where as our
sample is from a larger database (Blanton et al. 2002; {\tt
sample10}). For these galaxies, we use the use the $z$-band absolute
magnitude to estimate the mass as there is a nearly linear
relationship between this and the logarithm of the estimated stellar
mass in K03. This relationship is shown in Figure \ref{fig:fit}.  We
can see that the scatter is well restricted to a narrow range about
the line of best fit ($\rm
{Log_{10}(mass/M_{\odot})}=-0.5134M_{z}-0.01581$).  Using this fit we
estimate we can measure typical stellar masses to within a factor of
5.

\begin{figure}
\begin{centering}
\begin{tabular}{c}
{\epsfxsize=11truecm \epsfysize=11truecm \epsfbox[35 170 550 675
]{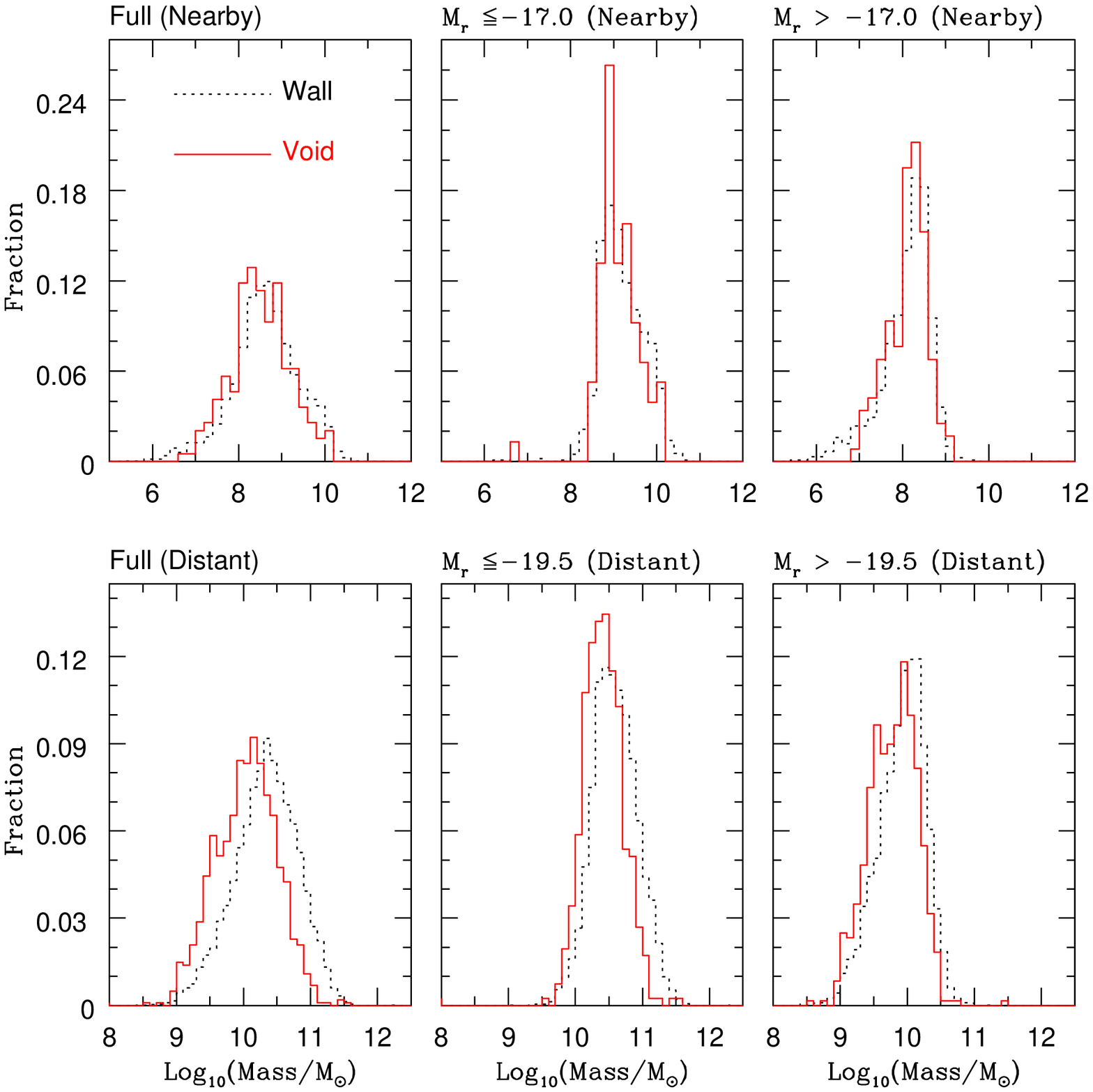}}
\end{tabular}
\caption{Stellar Mass Distribution. We show the normalized fraction of
void (solid lines) and wall galaxies (dotted lines) as a function of
$\rm{Log_{10}(Mass/M_{\odot})}$. The top row shows the results for the
nearby ($r<72 \mpc$) galaxies, the bottom row shows the results for
the distant ($100\leq r \leq260 \mpc$) galaxies.  The first, second
and third columns are the full, bright and faint samples. The fraction
of galaxies per $5M_{\odot}$ (nearby) and $10M_{\odot}$ (distant) bin
of $\rm{Log_{10}(Mass/M_{\odot})}$ is shown on the {\rm Y}-axis.  The
KS statistic reveals that the distant void galaxy (bright, faint, and
full) and respective wall galaxy samples are very different from one
another, with a probability of $<0.01\%$ that they are drawn from the
same parent population. In the case of the nearby galaxies, only the
bright galaxy distributions have a higher probability ($P\lsim0.36$)
of being similar.}
\label{fig:mass}
\end{centering}
\end{figure}

In Figure \ref{fig:mass} we show the normalized histograms of the
masses for all the samples. The differences between these
distributions are listed in Tables 1 and 2. It can be readily seen in
the distant sample that on average void galaxies have smaller stellar
masses than wall galaxies (10$^{10.13\pm0.02}M_{\odot}$ as opposed to
10$^{10.43}M_{\odot}$). The KS test shows that it is very unlikely
that void galaxies are drawn from the same parent population
($P<10^{-4}$). However, in the nearby samples the mean values of the
void and wall galaxy masses are more similar
(10$^{8.69\pm0.06}M_{\odot}$ and 10$^{8.80}M_{\odot}$).

\subsection{Star Formation Rates}
\label{sec:sfr}

In order to calculate SFR, the basic steps are 1.) measure the flux
in each line, 2.) convert the flux into a luminosity and 3.) use
Kennicutt's (1998) formulas for SFRs. Each of these steps is more
involved than it sounds. In general, we employ the SDSS SFR
prescription of Hopkins et al. (2003) with minor variations as
discussed below. We calculate the SFR for H$\alpha$ and [OII].

The flux is obtained via
\begin{equation}
F=\sqrt{2\pi}\sigma \kappa \times 10^{-17} \,[\rm{erg\,\, cm^{-2}\,\,
s^{-1}}]
\end{equation}
where $\sigma$ and $\kappa$ are the widths and heights of the Gaussian
fit to the respective line. These come from the SDSS spectroscopic
data analysis pipeline. $\kappa$ is in units of $[\rm{erg\,\,
cm^{-2}\,s^{-1}\, \AA}]$ and $\sigma$ in units of [{\AA}]. 

The next step is to to compute the respective luminosity ($L$) using:
\begin{equation}
L=4 \pi D_{L}^{2} F \,[\rm {erg\,\, s^{-1}}]
\end{equation}
where $D_{L}$ is the luminosity distance in units of [cm]. 

Finally, using Kennicutt's (1998b) formula for the H$\alpha$ flux
valid in the Case B recombination (Osterbrock 1986) with a Salpeter
(1955) initial mass function, we compute the SFR using
\begin{equation}
{\rm SFR}=7.9\times10^{-42} L\, [M_{\odot}\,\, yr^{-1}].
\end{equation}
and for [OII] we use Kennicutt's (1998) calibration
\begin{equation}
{\rm SFR}=1.4\times10^{-41} L\, [M_{\odot}\,\, yr^{-1}].
\end{equation}

However, each of these steps is more complicated than it seems as
there are several corrections that must be accounted
for. These include stellar absorption corrections, aperture effects, dust and
smearing. We discuss each of these below. 

Following Hopkins et al. (2003) we begin by applying the stellar
absorption correction. This consists of correcting the observed Balmer
line equivalent width (EW) by a factor (EW$_{c}$) appropriate for the
type of galaxy being studied. Hopkins et al. (2003) adopt a value of
EW$_{c}=1.3$ {\AA} for their analysis of the SDSS DR1 (Abazajian et
al. 2003) sample. Miller \& Owen (2002) found that this correction
varies from $0.9${\AA} (E type) to $4.1${\AA} (extreme spirals). In
our case a correction of EW$_{c}=1.4$ {\AA} is more appropriate for
the morphological mix of our galaxies (C. Tremonti private
communication). This correction is then converted to a flux correction
via $F_{c}=\frac{\rm {EW+EW_{c}}}{\rm EW}F$, where $F$ is the observed
line flux (Equation 1) and $F_{c}$ the corrected line flux.

The next step is to correct the stellar absorption corrected flux
$F_{c}$ for aperture effects. Because emission detected through the
$3''$ fiber from a galaxy depends greatly on the redshift of the galaxy
(nearby a smaller area of the galaxy is detected and at greater
distances a larger area can be detected) we need to estimate the flux
that would be measured in the $3''$ fiber. This is accomplished by
scaling $F_{c}$ by $10^{0.4(\rm{M_{r(spec)}}-\rm{M_{r(petro)}})}$
where M$_{\rm r(spec)}$ and M$_{\rm r(petro)}$ are the synthetic and
Petrosian\footnote{The magnitudes that are measured from the SDSS are
Petrosian magnitudes (Petrosian 1976).  The Petrosian magnitude is the
total amount of flux within a circular aperture whose radius depends
on the shape of the galaxy light profile, i.e., the size of the
aperture is not fixed so galaxies of the same type are observed out to
the same physical distance at all redshifts. More details are given in
Stoughton et al. (2002).} k-corrected SDSS $r$-band absolute
magnitudes respectively. Therefore, the stellar absorption and
aperture corrected flux ($F_{c}^{\prime}$) is
\begin{equation}
F_{c}^{\prime} = F_{c} \times 10^{0.4(\rm{M_{r(spec)}}-\rm{M_{r(petro)}})}  \,[\rm {erg\,\, s^{-1}}].
\end{equation}

By far the largest correction (about a factor of 10 
consists of accounting for obscuration by
dust. Here we apply the Fitzpatrick (1998) parameterization and
compute the color excess E(B-V) from the Balmer decrement given by
$\tau_{\rm Balmer}=\ln[\frac{F_{c}^{\prime}(\rm
H\alpha)}{F_{c}^{\prime}(\rm H\beta)}/2.87]$ where
$F_{c}^{\prime}(H\alpha)$ and $F_{c}^{\prime}(H\beta)$ are the stellar
absorption and aperture corrected H$\alpha$ and H$\beta$ fluxes
respectively and:
\begin{equation}
\rm
E(B-V)=1.086\, \tau_{\rm Balmer}/1.17. 
\end{equation}
Significant underestimates of the derived SFRs would be expected if
this effect were unaccounted for. In the case of SFRs derived from the
[OII] line, the situation is even more severe because of the larger
variation of $\tau_{\rm Balmer}$ (Gallagher et al. 1989; Kennicutt
1992a). This implies that SFRs of `dustier' galaxies will have
considerably larger corrections (increase in flux) than dust poor
galaxies. We find that void galaxies are relatively dust poor and
therefore, wall galaxies have the largest increase in flux due to
reddening corrections. This corrections are quite significant: wall
galaxies have an average factor of 13 increase in flux vs. void
galaxies which have a factor of 9 increase after all the relevant
corrections are applied.

Note that for $F_{c}^{\prime}(\rm H\beta)$ we use the $g$-band
absolute magnitude when estimating the aperture corrections. For all
other fluxes the $r$-band is used. With an estimate of the color
excess for each galaxy we can now de-redden the H$\alpha$, H$\beta$ and
[OII] fluxes using the R-dependent Galactic extinction curve of
Fitzpatrick \& Massa (1999). The dust corrected $F_{c}^{\prime}$ is
now substituted into Equation 2. to compute the respective absorption,
aperture, and dust corrected luminosity from which the SFR is derived
using Equation 3. 

Our analysis also accounts for smearing in the SDSS spectroscopic
observing procedure (Stoughton et al. 2002). It consists of four
minute spectroscopic observations performed while the telescope
dithers for the purpose of correcting for light losses from the $3''$
fiber aperture. For galaxies about 1/3 of their light is contained
within the $3''$ fiber aperture. Although the systematic differences
between smear corrected and original spectra are only about $10\%$
(Hopkins et al. 2003) the increase in the reported flux is generally
about a factor of 2. Fortunately this was accounted for by using the
$r$-band Petrosian magnitude in the aperture correction factor (this
essentially scales the spectrum to match the photometry which is more
reliable). An inherent assumption is that the EW(H$\alpha$) and
reddening are constant across the whole galaxy.

\begin{figure}
\begin{centering}
\begin{tabular}{c}
{\epsfxsize=11truecm \epsfysize=11truecm \epsfbox[35 170 550 675 ]{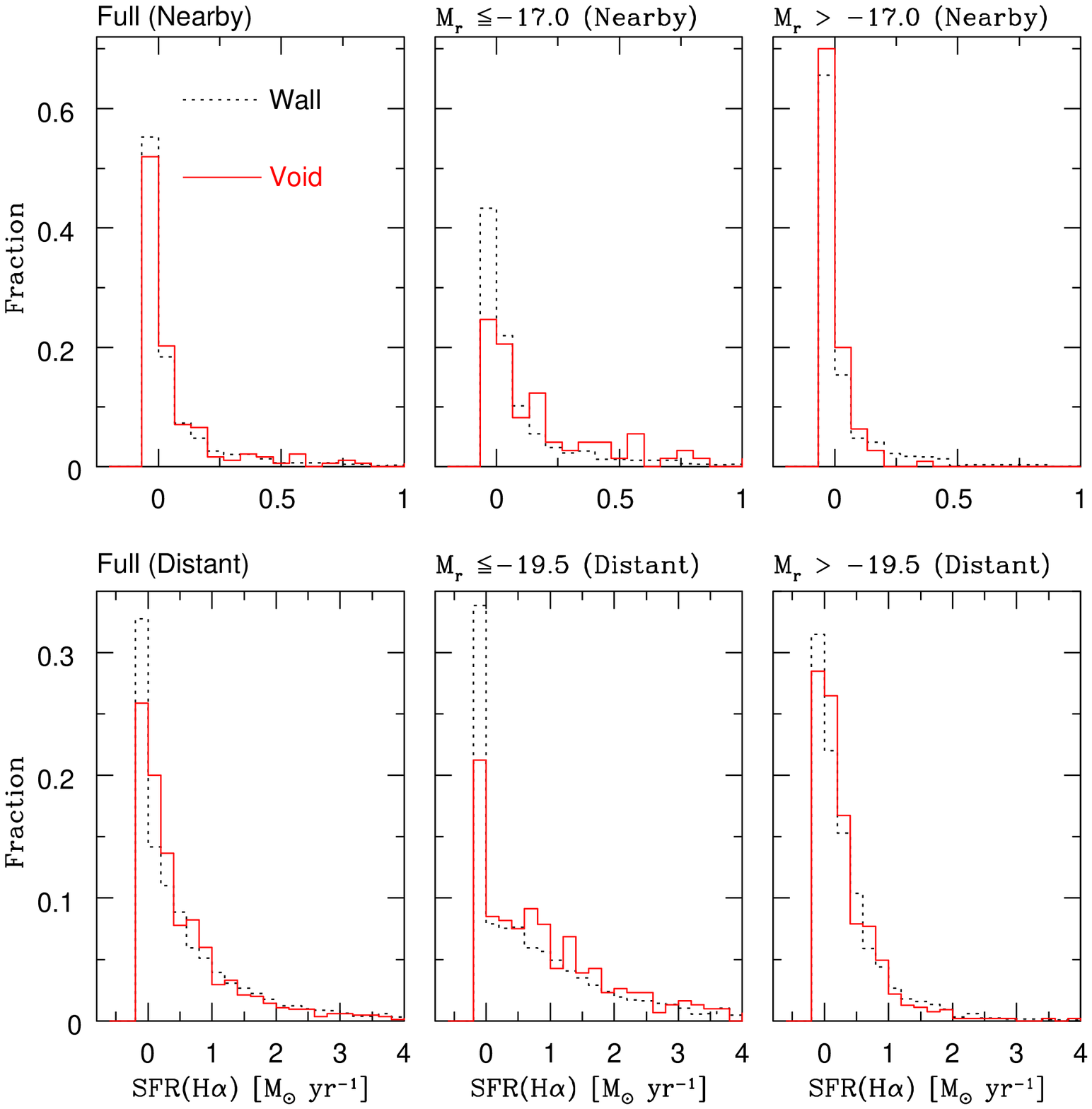}}
\end{tabular}
\caption{Distribution of H$\alpha$ star formation rates. We show the
normalized fraction of void (solid lines) and wall galaxies (dotted
lines) as a function of SFR(H$\alpha$). The top row shows the results
for the nearby ($r<72 \mpc$) galaxies, the bottom row shows the
results for the distant ($100\leq r \leq260 \mpc$) galaxies.  The
first, second and third columns are the full, bright and faint
samples. The fraction of galaxies per $0.067M_{\odot}\,\rm{ yr^{-1}}$
bin of SFR(H$\alpha$) is shown on the {\rm Y}-axis.  The KS statistic
reveals that the distant void galaxy (bright and full) and respective
wall galaxy samples are very different from one another, with a
probability of $<0.01\%$ that they are drawn from the same parent
population.}
\label{fig:Hasfr}
\end{centering}
\end{figure}

\begin{figure}
\begin{centering}
\begin{tabular}{c}
{\epsfxsize=11truecm \epsfysize=11truecm \epsfbox[35 170 550 675 ]{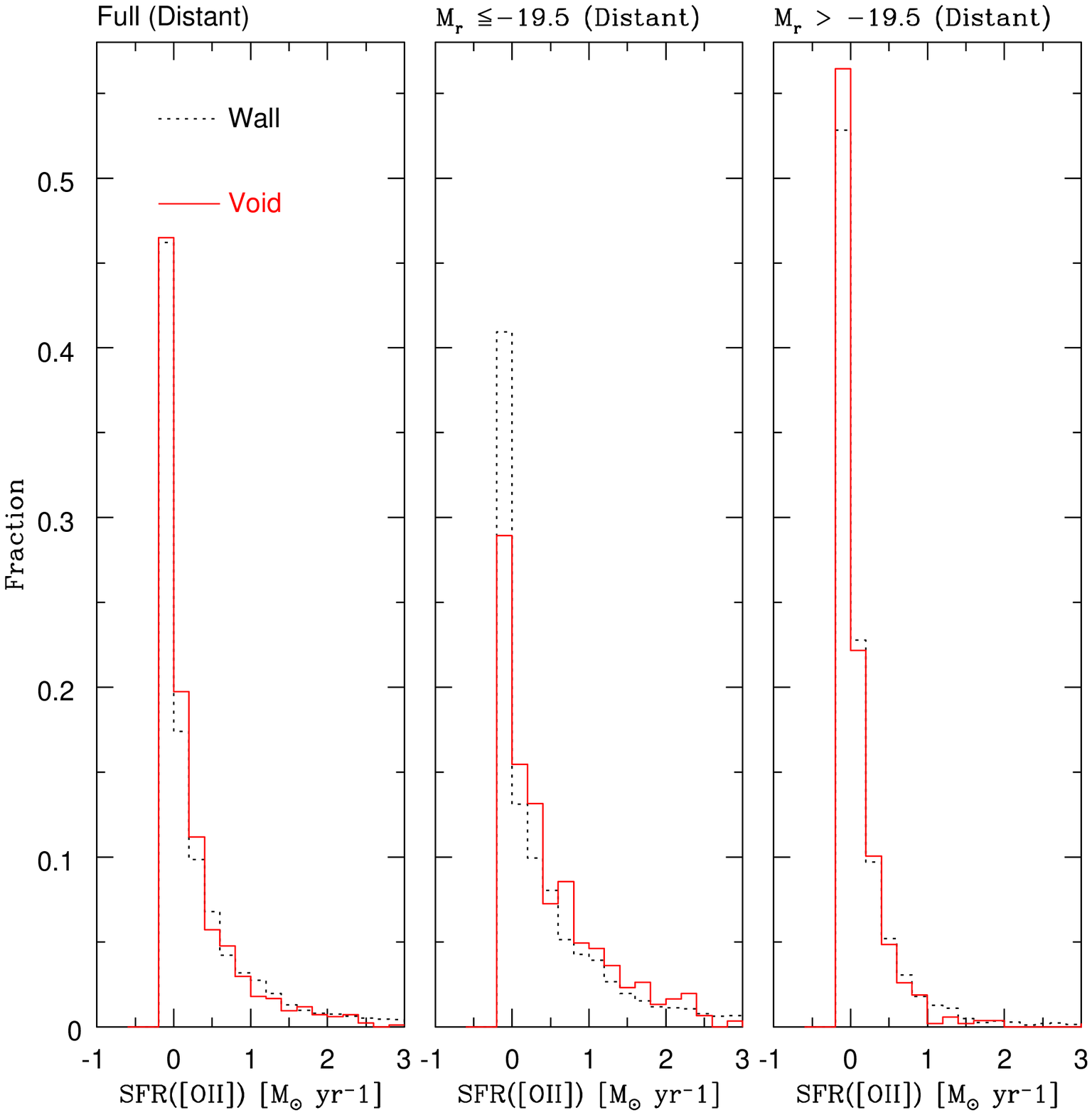}}
\end{tabular}
\caption{Distribution of [OII] star formation rates. We show the
normalized fraction of void (solid lines) and wall galaxies (dotted
lines) as a function of SFR([OII]).  The first, second and third
columns are the full, bright and faint distant samples. The fraction
of galaxies per $0.067M_{\odot}\,\rm{ yr^{-1}}$ bin of SFR([OII]) is
shown on the {\rm Y}-axis.  The KS statistic reveals that the distant
void galaxy (bright and full) and respective wall galaxy samples are
very different from one another, with a probability of $<0.01\%$ that
they are drawn from the same parent population.}
\label{fig:o2sfr}
\end{centering}
\end{figure}

In Tables 1 and 2 we summarize the results from the SFR estimates and
compare the SFR normalized distributions for void and wall galaxies in
Figures \ref{fig:Hasfr} (SFR(H${\alpha}$)) and \ref{fig:o2sfr}
(SFR([OII])). As noted above, SFR([OII]) can only be determined from
the distant sample as [OII] does not appear in the spectra of most of
the nearby sample of galaxies.

In the distant sample, the wall galaxies have slightly higher SFR than
the void galaxies (0.747$\pm0.007 M_{\odot}/yr$ vs.  0.734$\pm0.025
M_{\odot}/yr$ for H$\alpha$ and 0.488$\pm0.006 M_{\odot}/yr$
vs. 0.448$\pm0.015 M_{\odot}/yr$ for [OII]). However, in the nearby
sample, the void galaxies have slightly higher SFR than the wall
galaxies. Note carefully that comparing SFRs is a little misleading
though as the void galaxies are known to be fainter than the wall
galaxies and have smaller masses (section \ref{sec:mass}) and thus
would not be expected to have as high SFRs. A fairer comparison is to
use the specific star formation rate, which we discuss below.

Note that when computing the means, errors on the means, and the KS
statistic we include the negative SFRs in all the
calculations. Negative SFRs are obtained when either emission lines
are seen as absorption lines or when the strength of the line is too
weak and therefore, dominated by noise. The only SFRs that are not
included are those for which an estimate of the color excess was not
possible because $F_{c}^{\prime}(\rm H\alpha)/F_{c}^{\prime}(\rm
H\beta)<0$. Our samples of void and wall galaxies are reduced by about
$12\%$ due to this restriction.

\subsection{Specific Star Formation Rates}

The SFR per unit stellar mass (S-SFR) is perhaps a more informative
spectral indicator for our purposes because void galaxies are on
average less luminous and have smaller stellar masses. We estimate the
S-SFRs using the stellar masses and the absorption, aperture, and dust
corrected SFRs as discussed above. Note that as above, we have the
same caveat in our calculations, namely that the dust correction is a
large effect. Consistent with our analysis of the SFRs, we include
negative S-SFRs in the calculation of the mean. Neither the negative
SFRs or negative S-SFR have a physical meaning, yet if we remove the
$11\%$ of galaxies that have S-SFRs$<0$ the results would have an
artificial shift in the S-SFR distributions.

\begin{figure}
\begin{centering}
\begin{tabular}{c}
{\epsfxsize=11truecm \epsfysize=11truecm \epsfbox[35 170 550 675 ]{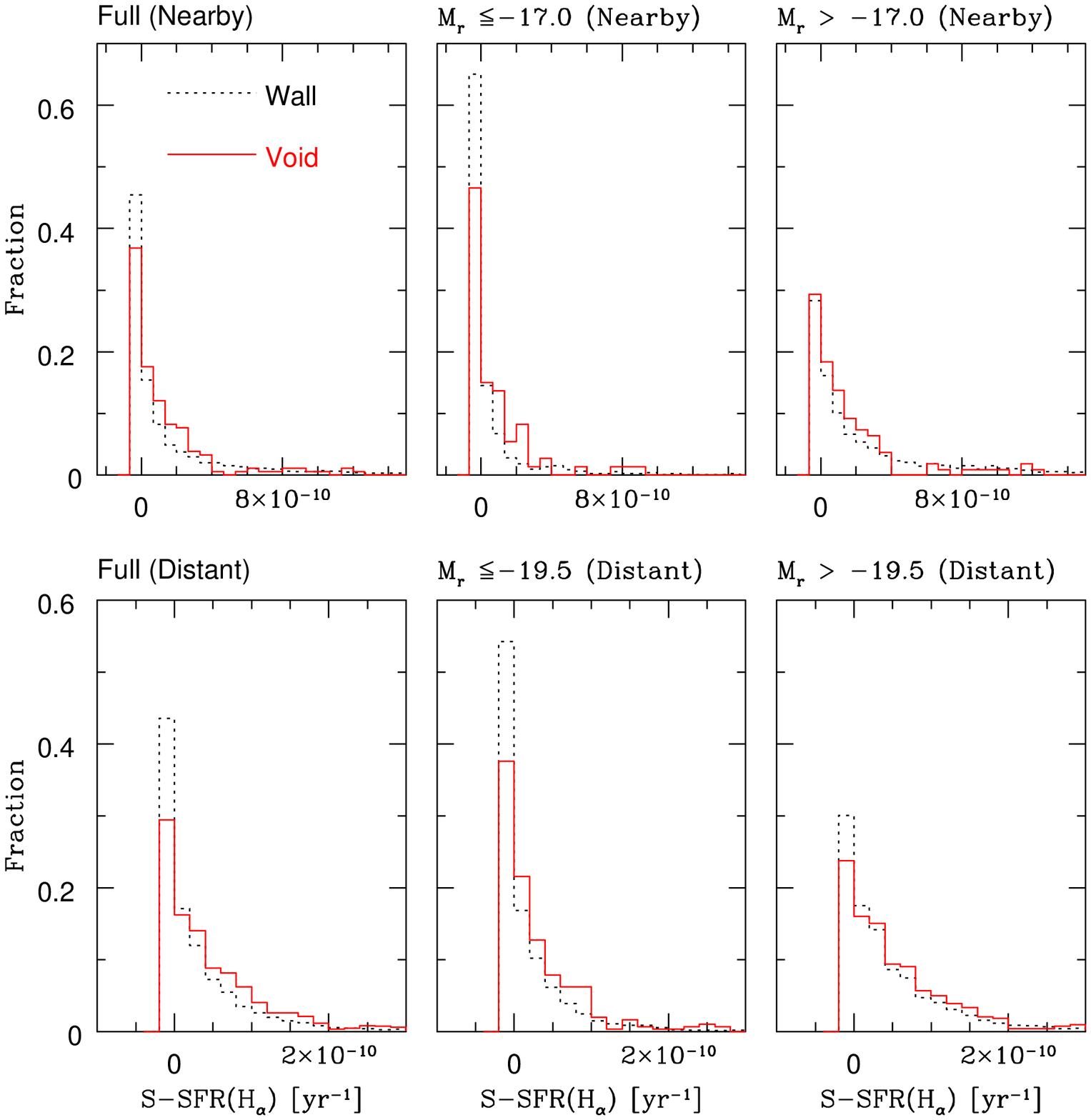}}
\end{tabular}
\caption{Distribution H$\alpha$ specific star formation rates. We show
the normalized fraction of void (solid lines) and wall galaxies
(dotted lines) as a function of S-SFR(H$\alpha$). The top row shows
the results for the nearby ($r<72 \mpc$) galaxies, the bottom row
shows the results for the distant ($100\leq r \leq260 \mpc$) galaxies.
The first, second and third columns are the full, bright and faint
samples. The fraction of galaxies per $10^{-10}\rm{ yr^{-1}}$ bin of
S-SFR(H$\alpha$) is shown on the {\rm Y}-axis.  The KS statistic
reveals that the distant void galaxy (bright, faint, and full) and
respective wall galaxy samples are very different from one another,
with a probability of $<0.01\%$ that they are drawn from the same
parent population. In the case of the nearby galaxies, only the bright
galaxy distributions have a higher probability ($P\lsim0.027$) of
being similar.}
\label{fig:Hassfr}
\end{centering}
\end{figure}

\begin{figure}
\begin{centering}
\begin{tabular}{c}
{\epsfxsize=11truecm \epsfysize=11truecm \epsfbox[35 170 550 675 ]{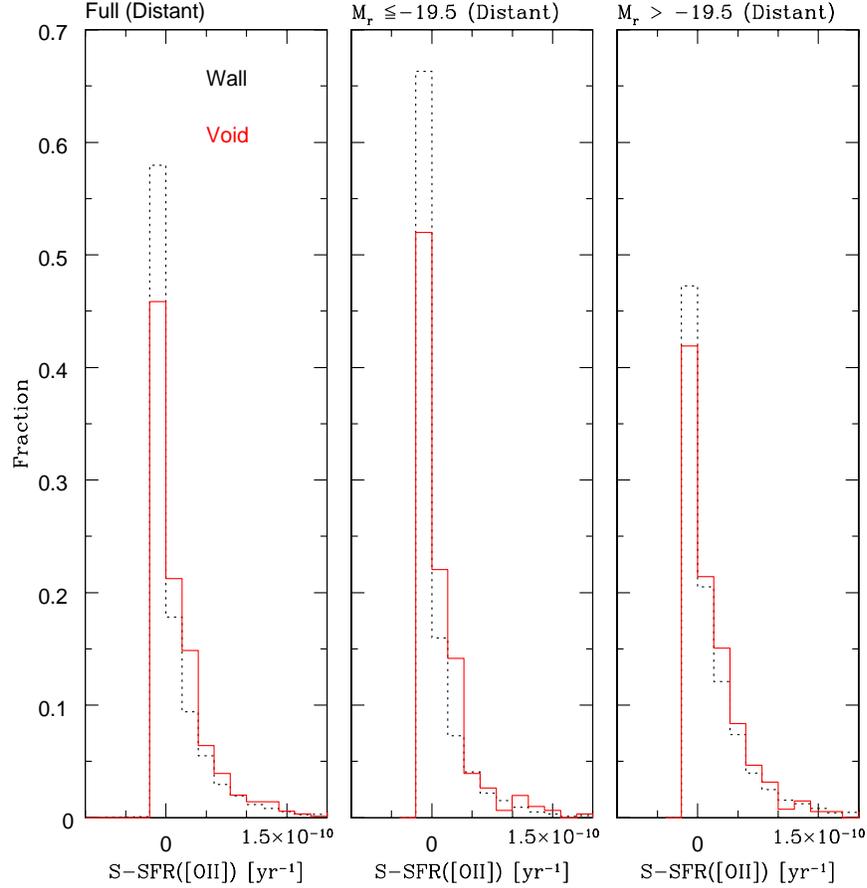}}
\end{tabular}
\caption{Distribution of [OII] specific star formation rates. We show
the normalized fraction of void (solid lines) and wall galaxies
(dotted lines) as a function of S-SFR([OII]).  The first, second and
third columns are the full, bright and faint distant samples. The
fraction of galaxies per $10^{-10} {\rm yr^{-1}}$bin of SFR([OII]) is
shown on the {\rm Y}-axis.  The KS statistic reveals that the distant
void galaxy (bright, faint, and full) and respective wall galaxy
samples are very different from one another, with a probability of
$<0.034\%$ that they are drawn from the same parent population.}
\label{fig:o2ssfr}
\end{centering}
\end{figure}

In Figures \ref{fig:Hassfr} and \ref{fig:o2ssfr} we show the S-SFR
distributions for the H${\alpha}$ and [OII] lines respectively. In the
distant samples (bright and faint), including sub-samples split by
S\'ersic index, void galaxies have larger average S-SFRs(H${\alpha}$)
and S-SFRs([OII]) than wall galaxies. The KS test shows that the
probability of void and wall galaxies being derived from the same
parent population is $<10^{-4}$. Thus, consistent with the results
from the EWs, void galaxies are forming more stars per unit mass than
the wall galaxies.

An apparent exception to this picture is that wall galaxies in the faint
nearby sample have higher SFRs and SSFRs than the void galaxies.
However, the EWs are the other way round; void galaxies have stronger
equivalent widths, although the differences are within
1-2$\sigma$. The photometric properties of these galaxies are also
least distinct, the void galaxies are slightly bluer but the
morphological properties are very similar (Rojas et al. 2004). It is
quite possible that the dust corrections applied to obtain the SFR are
not applicable for this class of faint, $M_r<-17$, galaxies or it is
possible that the wall galaxies in this category are also somewhat
isolated. A detailed study of the individual galaxies and/or a larger
sample that can be more finely split as a function of density is
required to study this ground of galaxies in more detail.

\section{Discussion}
\label{sec:disc}

To recap, we find that in all cases, distant or nearby, bright or
faint, that the void galaxies have higher EWs than galaxies that reside
in higher density regions. In most cases void galaxies also have
higher SSFRs than wall galaxies. In any case, equivalent width
measurements are less affected by the large uncertainties than the SFR
and SSFR calculations and, as such, are a more reliable indicator of
the recent star formation histories of galaxies. In this section, we
see if these results are consistent with work from other groups.

The role of a galaxy's environment on star formation has long been
studied. Recent work has tended to focus on galaxies in environments
that are more dense than the void galaxy sample that we present here.
The exception is the work of Grogin and Geller (2000), who probed the
SFR-density relation using a sample of 150 galaxies with values of
$n/\bar{n} <1$, i.e. $\delta \rho / \rho < 0$. 46 of these galaxies
had $\delta \rho / \rho < -0.5 $. They found that the strength of
EW(H$\alpha$) was 2-3$\sigma$ stronger than that of galaxies at higher
environments, consistent with our results. We are able to probe to
even lower density environments and find that the trend continues even
more strongly.

Balogh et al. (2004) also studied the strength of EW(H$\alpha$) in a
different range of environments, from groups to low density regions.
They consider both projected and three-dimensional density estimates,
the latter on scales of 1.1$h^{-1}$Mpc. They also find that the
strength of EW(H$\alpha$) increases as density decreases. They find
that rather than the whole population increasing in H$\alpha$, it is
the fraction of galaxies that show measurable H$\alpha$ that
dominates this increase. Even so, in their most underdense
environments, 30\% of the galaxies still show very little,
EW(H$\alpha<4$), emission. We find that 12-18\% of the void galaxies
have very little H$\alpha$ emission. 
They also
consider the density on scales of 5.5$h^{-1}$Mpc but due to a smaller
data set, they examined the SDSS EDR, the results were limited. They
suggested that on these larger scales, the difference between
EW(H$\alpha$) in void and wall regions might be greater, consistent
with our results.

G\'omez et al. (2003) go one step further and calculate SFRs and SSFRs
as a function of density. They use a projected density estimate so a
1-1 comparison is difficult and they do not probe environments that
are quite as underdense as those in this study. For example, the
strongest H$\alpha$ emitting galaxies have EW(H$\alpha$) of 19, which
is the average value for our void sample. Nevertheless, they also find
that SFR increases as the project density decreases, although there is
a large scatter. Lewis et al. (2004) perform a similar analysis using
the 2 degree Field Galaxy Redshift Survey and find very similar
results. G\'omez et al also examine whether the morphology density
relation is the reason for this change in star formation. They find
that late-type galaxies, as classified by the concentration index,
have stronger SFR in lower density environments than in high density
environments, consistent with our findings and results from Hashimoto
et al. (1998), Couch et al. (2001), and Pimbblet et al. (2001). 
Kauffmann et al. (2004) also probe SFR as a function of density,
although their lowest density bin includes more than 20\% of galaxies,
where as the void galaxy sample here includes only the lowest 5-8\% of
galaxies. The see similar trends of SFR with density, high star
formation in the lowest density regions, but in contrast to Balogh et
al. (2004) and the results here, they claim that the trend in SFR only
occurs on small, $<2 h^{-1}$Mpc, scales.

The observed photometric and spectroscopic properties of void galaxies
generally agree with predictions from semi-analytic models of
structure formation (Benson et al. 2003) and scenarios where
galaxy-galaxy interactions govern galaxy evolution. These models
attempt to include feedback effects from star formation. Void galaxies
modeled in this fashion are bluer, more disk-like, and have higher
specific star-formation rates than objects in denser regions. Benson
et al. (2003) find that the differences between void and wall galaxies
in their simulations are due to the shift toward smaller mass of the
halo mass function in voids. Evidence for such a shift in the halo
mass function is given by Goldberg et al. (2004), who estimate the
halo mass function of our void galaxy sample and find that the
best-fit theoretical mass function has mass underdensity
$\delta=-0.6$, similar to the underdensity in galaxies.

How do these results fit into the galaxy formation scenario? One
possibility is that void galaxies have formed recently and are young,
hence the young stellar ages, low masses and the active star formation
However, for galaxies to form in low density environments at all, they
would have to form before the matter in the voids flowed out to the
higher density regions. The relative blueness and active star formation
of the void galaxies would be explained if the void galaxies
consumed their gas supply at a slower rate than galaxies at higher density.

\section{Conclusions}
\label{sec:conc}

Using a sample of $\sim10^{3}$ void galaxies from the SDSS with
density contrast $\delta \rho / \rho < -0.6$, we are able to examine
the environmental dependence of spectral properties of galaxies in
extremely underdense regions.

We compare the equivalent widths of void and wall galaxies. Using this
simple approach, we find that void galaxies have larger equivalent
widths of H$\alpha$, [OII], H$\beta$, NII and [OIII]. This is the case
regardless of distance to the galaxy, brightness and morphological
type. This would suggest that void galaxies have undergone recent star
formation. 

Void galaxies also show weaker Balmer decrements, as measured by
D$_{n}(4000)$, which indicates that the stellar population includes a
higher fraction of O and B stars. Thus, the emission line widths and
continuum shape yield evidence of enhanced star formation over a range
of timescales.

Specific star-formation rates (estimated SFRs normalized by the
stellar mass of the galaxy) of void galaxies are generally higher than
for wall galaxies.  Un-normalized SFRs are similar or slightly higher
for wall galaxies than void galaxies, but the void galaxies tend to
have smaller stellar masses.

These spectroscopic results corroborate the photometric evidence given
by Rojas et al. (2004) that void galaxies are bluer than wall galaxies
of similar luminosity and morphology. 

Thus, we show that the trend of increasing SFR with decreasing
density, seen heretofore in denser regions, does indeed continue into
the rarefied environment of voids.

\section*{Acknowledgments}

Funding for the creation and distribution of the SDSS Archive has been
provided by the Alfred P. Sloan Foundation, the Participating
Institutions, the National Aeronautics and Space Administration, the
National Science Foundation, the U.S. Department of Energy, the
Japanese Monbukagakusho, and the Max Planck Society. The SDSS Web site
is http://www.sdss.org/.

The SDSS is managed by the Astrophysical Research Consortium (ARC) for
the Participating Institutions. The Participating Institutions are The
University of Chicago, Fermilab, the Institute for Advanced Study, the
Japan Participation Group, The Johns Hopkins University, Los Alamos
National Laboratory, the Max-Planck-Institute for Astronomy (MPIA),
the Max-Planck-Institute for Astrophysics (MPA), New Mexico State
University, University of Pittsburgh, Princeton University, the United
States Naval Observatory, and the University of Washington.

We thank Christy Tremonti and Andrew Hopkins for useful
conversations and for assistance with the systematics in the SFR
calculations. MSV acknowledges support from NSF grant AST-0071201 and
a grant from the John Templeton Foundation.

\begin{table}
\tiny{
\begin{centering}
\begin{tabular}{c}
\bf{DISTANT SAMPLE}
\end{tabular}
\begin{tabular}{cccccccc}

& &  \bf {Full} ($-22.5 \le M_{r} \le -17.77$)   & [$N_{V}=1010$, $N_{W}=12732$] &\\ \hline
& & Void (VGD)&  Wall (WGD) &  KS ($P$) & [$N_{V}, N_{W}$] \\ 
Property &  &$\mu\pm\sigma_{\mu}$  & $\mu\pm\sigma_{\mu}$ & Probability  &\\ \hline

EW($\rm{H}\alpha$)\,[\AA] & & $19.14\pm0.680$ & $11.77\pm0.158$ & $<10^{-4}$ & (1005, 12636) &\\
EW([OII])\,[\AA]      & & $14.26\pm0.395$ & $9.441\pm0.093$ & $<10^{-4}$ & (1006, 12301) &\\ 
EW($\rm {H}\beta$)\,[\AA]  & & $2.571\pm0.146$  & $0.828\pm0.035$ & $<10^{-4}$ & (997, 12544) &\\
EW([NII])\,[\AA]     & & $6.553\pm0.181$  & $5.065\pm0.049$ & $<10^{-4}$ & (1006, 12586) &\\ 
EW([OIII])\,[\AA]    & & $4.694\pm0.335$  & $2.197\pm0.061$ & $<10^{-4}$ & (994, 12373) &\\ 
$D_{n}4000$  & & $1.515\pm0.011$  & $1.649\pm0.003$ & $<10^{-4}$ & (714, 9862) &\\ 
${\rm Log}_{10}({\rm Mass}/M_{\odot})$  & & $10.13\pm0.015$  & $10.43\pm0.004$ & $<10^{-4}$ & (1010, 12732) &\\ 
SFR($\rm{H}\alpha$) $[M_{\odot}/yr]$  & & $0.734\pm0.025$  & $0.747\pm0.007$ & $<10^{-4}$ & (850, 8930) &\\
SFR($[OII]$) $[M_{\odot}/yr]$ & & $0.448\pm0.015$  & $0.488\pm0.006$ & $<10^{-4}$ & (841, 8764) &\\
S-SFR($\rm {H}\alpha$) $[yr^{-1}]$ & & $(3.744\pm0.108)\times10^{-11}$  & $(2.629\pm0.034)\times10^{-11}$ & $<10^{-4}$ & (850, 8930) &\\
S-SFR([OII]) $[yr^{-1}]$ & & $(4.892\pm0.890)\times10^{-11}$  & $(3.020\pm0.251)\times10^{-11}$ & $<10^{-4}$ & (841, 8764) &\\ \hline \\

& &  \bf {Bright} ($M_{r} \le -19.5$)  &[$N_{V}=409$, $N_{W}=7831$] &\\ \hline
& & Void (VGD\_b)&  Wall (WGD\_b) &  KS ($P$) &  [$N_{V}, N_{W}$] \\ 
Property &  &$\mu\pm\sigma_{\mu}$  & $\mu\pm\sigma_{\mu}$ & Probability  &\\ \hline
EW($\rm{H}\alpha$)\,[\AA] & & $14.47\pm0.932$ & $8.542\pm0.175$ & $<10^{-4}$ & (406, 7774) &\\
EW([OII])\,[\AA]       & & $9.338\pm0.429$  & $6.884\pm0.089$ & $<10^{-4}$ & (399, 7548) &\\ 
EW($\rm{H}\beta$)\,[\AA]  & & $1.246\pm0.201$  & $0.018\pm0.038$ & $<10^{-4}$ & (401, 7699) &\\
EW([NII])\,[\AA]       & & $6.606\pm0.332$  & $4.711\pm0.066$ & $<10^{-4}$ & (407, 7743) &\\ 
EW([OIII])\, [\AA]     & & $2.706\pm0.531$  & $1.172\pm0.055$ & $<10^{-4}$ & (401, 7574) &\\ 
$D_{n}4000$  & & $1.602\pm0.017$  & $1.711\pm0.004$ & $<10^{-4}$ & (287, 6069) &\\ 
${\rm Log}_{10}({\rm Mass}/M_{\odot})$  & & $10.52\pm0.016$  & $10.70\pm0.004$ & $<10^{-4}$ & (409, 7831) &\\ 
SFR($\rm{H}\alpha$) $[M_{\odot}/yr]$ & & $1.136\pm0.063$  & $0.920\pm0.016$ & $<10^{-4}$ & (306, 4956) &\\
SFR([OII]) $[M_{\odot}/yr]$ & & $0.736\pm0.051$  & $0.626\pm0.013$ & $<10^{-4}$ & (304, 4876) &\\
S-SFR($\rm{H}\alpha$) $[yr^{-1}]$ & & $(3.133\pm0.169)\times10^{-11}$  & $(2.137\pm0.004)\times10^{-11}$ & $<10^{-4}$ & (306, 4956) &\\
S-SFR([OII]) $[yr^{-1}]$ & & $(3.061\pm0.279)\times10^{-11}$  & $(2.192\pm0.053)\times10^{-11}$ & $<10^{-4}$ & (304, 4876) &\\ \hline \\

& &  \bf {Faint} ($M_{r}>-19.5$)  & [$N_{V}=601$, $N_{W}=4901$]&\\ \hline
& & Void (VGD\_f)&  Wall (WGD\_f) &  KS ($P$) &  [$N_{V}, N_{W}$] \\ 
Property &  &$\mu\pm\sigma_{\mu}$  & $\mu\pm\sigma_{\mu}$ & Probability  &\\ \hline

EW($\rm{H}\alpha$)\, [\AA] & & $22.31\pm0.929$ & $16.93\pm0.287$ & $<10^{-4}$ & (599, 4862) &\\
EW([OII])\,[\AA]      & & $17.57\pm0.554$ & $13.50\pm0.178$ & $<10^{-4}$ & (593, 4753) &\\ 
EW($H\beta$)\,[\AA]  & & $3.452\pm0.202$  & $2.095\pm0.067$ & $<10^{-4}$ & (531, 4334) &\\
EW([NII])\,[\AA]      & & $6.517\pm0.204$  & $5.630\pm0.074$ & $<10^{-4}$ & (599, 4843) &\\ 
EW([OIII])\,[\AA]     & & $6.038\pm0.423$  & $3.817\pm0.128$ & $<10^{-4}$ & (593, 4799) &\\ 
$D_{n}4000$  & & $1.456\pm0.013$  & $1.551\pm0.004$ & $<10^{-4}$ & (427, 3793) &\\ 
${\rm Log}_{10}({\rm Mass}/M_{\odot})$  & & $9.873\pm0.015$  & $10.02\pm0.005$ & $<10^{-4}$ & (601, 4901) &\\ 
SFR($\rm{H}\alpha$) $[M_{\odot}/yr]$ & & $0.508\pm0.0237$  & $0.530\pm0.009$ & $0.0254$ & (544, 3974) &\\
SFR([OII]) $[M_{\odot}/yr]$ & & $0.286\pm0.017$  & $0.316\pm0.007$ & $0.0298$ & (537, 3888) &\\
S-SFR($\rm{H}\alpha$) $[yr^{-1}]$  & & $(4.146\pm0.137)\times10^{-11}$  & $(3.349\pm0.005)\times10^{-11}$ & $<10^{-4}$ & (544, 3974) &\\
S-SFR([OII]) $[yr^{-1}]$  & & $(5.725\pm1.375)\times10^{-11}$  & $(4.058\pm0.147)\times10^{-11}$ & $0.0034$ & (537, 3888) &\\ \hline \\
\end{tabular}
\caption{Means, errors on the means and KS test probabilities that the
void and wall galaxies are drawn from the same parent population for
the spectroscopic properties of void and wall galaxies in the distant
sample ($100\leq r \leq260 \mpc$). The number of galaxies (void and
wall) in each sample and sub-sample are listed next to the magnitude
range heading as [$N_{V}$ (void), $N_{W}$ (wall)]. Small values of $P$
correspond to a low probability that the two samples are drawn from
the same parent population.  In this case, the KS test shows that the
void and wall galaxies are drawn from different populations based on
emission line EWs, stellar masses, H$\alpha$ and [OII] derived SFRs
and S-SFRs, and D$_{n}$4000. The differences between the means of the
different parameters measured are on average $> 5 \sigma_{\mu}$,
except for the SFR($\rm{H}\alpha$) and SFR([OII]) in the faint
sub-sample, where the difference is $\sim 2\sigma_{\mu}$. Void
galaxies on average have higher S-SFRs, larger EWs, smaller stellar
masses and smaller $D_{n}4000$.}
\label{tab:prop1}
\end{centering}
}
\end{table}

\begin{table}
\tiny{
\begin{centering}

\begin{tabular}{c}
\bf{NEARBY SAMPLE}
\end{tabular}

\begin{tabular}{cccccccc}
& &  \bf {Full} ($-19.9 \le M_{r} \le -14.5$)   & [$N_{V}=194$, $N_{W}=2256$] &\\ \hline
& & Void (VGD)&  Wall (WGD) &  KS ($P$) & [$N_{V}, N_{W}$] \\ 
Property &  &$\mu\pm\sigma_{\mu}$  & $\mu\pm\sigma_{\mu}$ & Probability  &\\ \hline

EW($\rm{H}\alpha$)\,[\AA] & & $35.31\pm0.262$ & $26.18\pm0.645$ & $0.0021$ & (187, 2202) &\\
EW($\rm{H}\beta$)\,[\AA]  & & $7.317\pm0.738$  & $5.702\pm0.242$ & $0.0007$ & (192, 2226) &\\
EW([NII])\,[\AA]      & & $5.175\pm0.338$  & $4.298\pm0.097$ & $0.0611$ & (191, 2232 ) &\\ 
EW([OIII])\,[\AA]     & & $18.50\pm2.111$  & $14.17\pm0.548$ & $<10^{-4}$ & (188, 2211) &\\ 
$D_{n}4000$  & & $1.261\pm0.019$  & $1.314\pm0.005$ & $<10^{-4}$ & (119, 1545) &\\ 
${\rm Log}_{10}({\rm Mass}/M_{\odot})$  & & $8.692\pm0.060$  & $8.803\pm0.018$ & $0.0592$ & (194, 2256) &\\ 
SFR($\rm{H}\alpha$) $[M_{\odot}/yr]$ & & $0.162\pm0.012$  & $0.146\pm0.006$ & $0.3410$ & (183, 2020) &\\
S-SFR($\rm{H}\alpha$) $[yr^{-1}]$ & & $(25.55\pm0.353)\times10^{-11}$  & $(31.70\pm0.169)\times10^{-11}$ & $0.0001$ & (182, 2005) &\\ \hline \\

& &  \bf {Bright} ($M_{r} \le -17.0$)  &[$N_{V}=76$, $N_{W}=1071$] &\\ \hline
& & Void (VGD\_b)&  Wall (WGD\_b) &  KS ($P$) &  [$N_{V}, N_{W}$] \\ 
Property &  &$\mu\pm\sigma_{\mu}$  & $\mu\pm\sigma_{\mu}$ & Probability  &\\ \hline
EW($\rm{H}\alpha$) \,[\AA] & & $33.316\pm3.74$ & $21.91\pm0.809$ & $0.0005$ & (76, 1060) &\\
EW($\rm{H}\beta$)\,[\AA]  & & $5.939\pm0.769$  & $3.770\pm0.210$ & $0.0006$ & (76, 1055) &\\
EW([NII])\, [\AA]      & & $7.223\pm0.733$  & $5.203\pm0.157$ & $0.0034$ & (76, 1062) &\\ 
EW([OIII])\,[\AA]     & & $11.615\pm2.41$  & $8.276\pm0.525$ & $0.0133$ & (76, 1056) &\\ 
$D_{n}4000$  & & $1.287\pm0.032$  & $1.371\pm0.008$ & $<10^{-4}$ & (50, 792) &\\ 
${\rm Log}_{10}({\rm Mass}/M_{\odot})$  & & $9.333\pm0.066$  & $9.390\pm0.018$ & $0.4123$ & (76, 1071) &\\ 
SFR($\rm{H}\alpha$) $[M_{\odot}/yr]$  & & $0.323\pm0.048$  & $0.194\pm0.011$ & $0.0009$ & (73, 933) &\\
S-SFR($\rm{H}\alpha$) $[yr^{-1}]$ & & $(17.57\pm3.222)\times10^{-11}$  & $(12.57\pm0.958)\times10^{-11}$ & $<10^{-4}$ & (73, 933) &\\ \hline \\

& &  \bf {Faint} ($M_{r}>-17.0$)  & [$N_{V}=118$, $N_{W}=1185$]&\\ \hline
& & Void (VGD\_f)&  Wall (WGD\_f) &  KS ($P$) &  [$N_{V}, N_{W}$] \\ 
Property &  &$\mu\pm\sigma_{\mu}$  & $\mu\pm\sigma_{\mu}$ & Probability  &\\ \hline

EW($\rm{H}\alpha$)\,[\AA] & & $36.68\pm3.60$ & $30.14\pm0.977$ & $0.3740$ & (111, 1142) &\\
EW($\rm{H}\beta$)\,[\AA]  & & $8.220\pm1.111$  & $7.443\pm0.412$ & $0.2920$ & (116, 1171) &\\
EW([NII])\,[\AA]      & & $3.821\pm0.378$  & $3.477\pm0.121$ & $0.4446$ & (115, 1170) &\\ 
EW([OIII])\,[\AA]     & & $23.17\pm3.073$  & $19.55\pm0.905$ & $0.0481$ & (112, 1155) &\\ 
$D_{n}4000$  & & $1.243\pm0.022$  & $1.253\pm0.007$ & $<10^{-4}$ & (69, 753) &\\ 
${\rm Log}_{10}({\rm Mass}/M_{\odot})$  & & $8.279\pm0.060$  & $8.273\pm0.019$ & $0.3589$ & (118, 1185) &\\ 
SFR($\rm{H}\alpha$) $[M_{\odot}/yr]$ & & $0.054\pm0.006$  & $0.098\pm0.005$ & $0.1725$ & (109, 1066) &\\
S-SFR($\rm{H}\alpha$) $[yr^{-1}]$ & & $(29.19\pm4.527)\times10^{-11}$  & $(46.37\pm2.713)\times10^{-11}$ & $0.0269$ & (109, 1066) &\\ \hline \\
\end{tabular}
\caption{Means, errors on the means and KS test probabilities that the
void and wall galaxies are drawn from the same parent population for
the spectroscopic properties of void and wall galaxies in the nearby
sample ($r<72 \mpc$). The number of galaxies (void and wall) in each
sample and sub-sample are listed next to the magnitude range heading
as [$N_{V}$ (void), $N_{W}$ (wall)]. Small values of $P$ correspond to
a low probability that the two samples are drawn from the same parent
population. The KS test shows that void galaxies appear to have
stronger emission line EWs than wall galaxies in all cases. The
average difference between the means of the EWs and $D_{n}4000$s is
about $2 \sigma_{\mu}$.  However, the SFRs and stellar masses are not
significantly different. Only the bright void galaxy
sub-sample shows a larger S-SFR($\rm{H}\alpha$) than wall
galaxies.}
\label{tab:prop2}
\end{centering}
}
\end{table}

\end{document}